\documentstyle[12pt]{article}
\begin{document}

\centerline{\bf Heating-Assisted Atom Transfer in the Scanning
Tunneling Microscope }  

\vspace{2cm}
\centerline{ M. Grigorescu\footnote{ present address: Institut 
f\"ur Theoretische Physik, Justus-Liebig-Universit\"at Giessen,
Heinrich-Buff-Ring 16, D-35392 Giessen, Germany } } 

\vspace{5cm}
{\bf Abstract}: \\ 
The effect of the environmental interactions on the localization
probability for a Xe atom  trapped in the 
surface-tip junction of the scanning tunneling microscope is studied in 
the frame of a stochastic, non-linear Liouville equation for the
density operator. It is shown that the irreversible transfer from
surface to tip may be explained  by thermal decoherence rather 
than by the driving force acting during the application of a
voltage pulse. 
 \\[1cm]
PACS numbers: 61.16.Di,73.40.Gk,05.40.+j  \\[1cm] 
Submitted to Z. Phys. B.  \\[.5cm]
http://publish.aps.org/eprint/gateway/eplist/aps1997aug19\_001

\newpage

{\bf 1. Introduction} \\[.5cm]
\indent
Since the first experiments on reversible atom 
transfer   \cite{ei}, the  scanning tunneling microscope (STM) may be 
considered as the ideal instrument for  manipulating atoms or molecules. 
The bistable operation mode of STM indicate that in a certain geometry
the diffusion barrier on surface is high enough to prevent
the particle escape from the junction region, and the motion takes
place along the outer normal to the surface plane  in an 
asymmetric, one-dimensional, double-well  potential (DWP) \cite{walk}. 
However, the mechanism of irreversible atom transfer between
the potential wells after the application of a voltage pulse
is not yet completely understood.
\\ \indent
If the initial center of mass (CM) wave function of the atom is 
an isomeric state of the static junction potential, then
the barrier crossing appears only in  special  resonance conditions, by  
quantum coherence  oscillations (QCO) \cite{gbc}.  The
resonances with long oscillation periods require a very fine tunning of 
the bias voltage, which cannot be assumed in the present switching 
experiments. Also, during QCO the tunneling has not an exponential
law. Therefore, is a most interesting issue to study the evolution
of the atomic CM wave packet dynamics  during the voltage pulse, and the 
decoherence effects produced by the coupling with the environment. 
This coupling may damp \cite{bend} or completely suppress \cite{bm} the QCO. 
\\ \indent
The appropriate frame for the treatment of a quantum dynamics
determined both by the unitary evolution in the Hilbert space
and by the change in the purity of the states (of the occupation numbers), 
is provided by the Liouville equation for the density operator. However, this 
equation can be solved only for very simple systems, and in a physical
situation is necessary to find suitable approximations.  \\ \indent
The occupation numbers of the energy levels  for a Xe atom in the STM 
junction potential may change in time by the interactions with the
electron gas. A partial description of this process is given by
the rate equations \cite{walk}, but for a realistic treatment is
necessary to account also the dynamical effects related to the 
evolution of the non-stationary wave packets, as the QCO. 
 \\ \indent  
In this work, the effect of environmental decoherence on the
atom dynamics in the STM junction will be described phenomenologically
using a modified Liouville equation. This equation is presented in Sec. 2,  
and is applied to calculate the transition rate induced by the thermal noise. 
In Sec. 3 are presented numerical results concerning 
the evolution of the localization probability for a Xe atom.
At zero temperature, the probability density during a voltage pulse is
determined by the evolution of the quantum wave packet in a time-dependent
external potential. This evolution is obtained
by solving numerically the  Schr\"odinger equation,
for symmetric triangular and trapezoidal pulses with $-0.8 $ V
at the peak. The environmental decoherence effects are studied using the
non-linear Liouville equation for the density operator,
at the temperature $T=4$ K and a constant voltage $U$. 
Two values of $U$ are considered, one non-resonant, 
$U=-0.8$ V, and  $U-1.141$ V, corresponding to the  QCO resonance with
the highest frequency. The conclusions are summarized in  Sec. 4.  
\\[.5cm]
{\bf 2. Brownian quantum dynamics}
\\[.5cm] \indent
Let us  consider a quantum system with the density operator
${\cal D}$, interacting with 
a classical heat bath of $N_c$ harmonic oscillators via the bilinear coupling 
term
\cite{hang}
\begin{equation}
H_{coup} = x \sum_{i=1}^{N_c} C_i q_i ~,
\end{equation}
where $x$ is the coupling operator, $C_i$ are constants and $q_i$ the 
time-dependent bath coordinates. 
The evolution of this mixed classical-quantum system can be
obtained from the variational equation \cite{mg2}
\begin{equation}
\delta \int dt \{ \sum_{i=1}^{N_c} (  \dot q_i p_i  -  h_i ) +
Tr \lbrack {\bf \eta}^\dagger i \hbar \partial_t {\bf \eta} - 
\eta^\dagger ({ H_0 }+ { H}_{coup}) {\bf \eta}  \rbrack \} =0~~,
\end{equation}
where $h_i = (p_i^2  + m_i^2  \omega^2_i q_i^2 )/2 m_i$ is the classical 
Hamiltonian for a bath oscillator, ${ H}_0$  is the Hamiltonian operator 
for  the isolated quantum system,
and $ \eta$, $\eta^\dagger$ are the "square root operators"  defined by the
Gauss decomposition of the density operator,  ${\cal D}= \eta \eta^\dagger$ 
\cite{reznik}.
In the physical situation of a thermally equilibrated bath with infinite
heat capacity, Eq. (2) leads to a Brownian dynamics of the density
operator ${\cal D}$, described by the stochastic Liouville equation 
\cite{mg2}
\begin{equation}
i \hbar \partial_t {\cal D} = [{ H}_0  
- x (\xi(t) + f_{\cal D} (t) ) , {\cal D}]~~.
\end{equation} 
Here $\xi(t)$ is a random force with zero mean (the noise), while 
\begin{equation}
f_{\cal D} (t) = - \int_0^t dt' \Gamma(t-t') \frac{ d Tr( {\cal D} x)}{dt'}
\end{equation}
is a friction force with the memory function $\Gamma(t)$. If $<<*>>$
denotes the average over the statistical bath ensemble at the temperature $T$,
then  $<<\xi(t)>>=0$, and $<<\xi(t) \xi(t')>>=k_BT \Gamma(t-t')$,
by the  fluctuation-dissipation theorem (FDT). 
The Brownian evolution of ${\cal D}$ determined by Eq. (3) preserves the 
"purity" of the initial state, but decoherence may appear for the average
\begin{equation}
{\cal D}_{av} (t)  \equiv << {\cal D}>> (t)= \sum_{r=1}^{N_t} 
\frac{ {\cal D}^r {(t)}}{N_t}
\end{equation}  
calculated over an ensemble of $N_t$ trajectories ${\cal D}^r {(t)}$ generated 
with the same initial condition, ${\cal D}^r (0) = {\cal D}_0$.
Of particular interest is the case when the initial state of the
system is not thermally equilibrated, and  ${\cal D}_0 = \mid \psi_0>< \psi_0 
\mid $,  with $\mid \psi_0>$ a pure state. In this case, 
\begin{equation}
{\cal D}_{av} (t) = \frac{1}{N_t} \sum_{r=1}^{N_t} 
 \mid \psi^r (t) >< \psi^r (t) \mid  
\end{equation}  
where $\mid \psi^r (t)>$ is a solution of the modified Schr\"odinger equation
\begin{equation}
i \hbar \partial_t \mid \psi^r > = [{ H}_0  
- x (\xi(t) + f_{\cal D}^r (t)) ] \mid \psi^r >~~.
\end{equation} 
\\ \indent
If the friction may be neglected, then Eq. (3)  has the general solution
\begin{equation}
{\cal D}^r(t) = e^{- i  H_0 t / \hbar} \tilde{ \cal D}^r (t) e^{i H_0 t /
\hbar}~~,
\end{equation}
with
\begin{equation}
\tilde{\cal D}^r {(t)} = {\cal T} e^{ \frac{i}{\hbar} \int_0^t dt' 
\xi^r (t') {\cal L}_{\tilde{x} (t')} }~ \rho_{0} ~~.
\end{equation}
Here ${\cal T}$ denotes the time-ordering operator \cite{bd}, 
${\cal L}_A$ is the Lie derivative with respect to the operator $A$
defined by the commutator, ${\cal L}_A B  \equiv [A,B]$, while
\begin{equation}
\tilde{x} (t) =  e^{ i  H_0 t / \hbar} x e^{-i H_0 t /
\hbar}
\end{equation}
 is the coupling operator in the interaction representation.  
\\ \indent
The evolution determined by the unitary operator $\exp(- i H_0 t
/ \hbar)$ is the same for all the trajectories appearing in Eq. (5),
and therefore the ensemble average may be written as 
\begin{equation}
{\cal D}_{av} (t)= e^{- i  H_0 t / \hbar} \tilde{ \cal D}_{av} (t)
e^{i H_0 t / \hbar}~~.
\end{equation}
Here $\tilde{ \cal D}_{av} (t)$ denotes the trajectory average of
$\tilde{\cal D}^r (t)$, and can be calculated using the FDT 
after the expansion of the time-ordered exponential in
Eq. (9). Retaining only the first non-vanishing average, the result is
\begin{equation} 
\tilde{ \cal D}_{av}(t) = {\cal D}_0 - \frac{ k_B T}{\hbar^2}
\int_0^t dt_1 \int_0^{t_1} dt_2 \Gamma(t_1-t_2) [ \tilde{x} (t_1),
[ \tilde{x} (t_2), {\cal D}_0 ] ] ~~.
\end{equation}
This formula can be applied to calculate the rate of the noise-induced 
transitions between the energy eigenstates $\{ \mid E_k> \}$ of
$H_0$, defined by  $H_0 \mid E_k> = E_k \mid E_k>$. 
If initially the system is in the pure state
$\mid E_i>$, then ${\cal D}_0 = \mid E_i><E_i \mid$, and the rate of the  
transition $\mid E_i> \rightarrow  \mid E_f>$ is given by the asymptotic
time-derivative 
\begin{equation}
\lambda_{fi} =
 \frac{d v_f}{dt} \mid_{t \rightarrow \infty}
\end{equation}
of the occupation probability
$
v_f (t)= < E_f \mid {\cal D}_{av} (t) \mid E_f>=  < E_f \mid
\tilde{ {\cal D}}_{av} (t) 
\mid E_f> $. Using Eq. (12), this probability is 
\begin{equation}
v_f (t) = 2  k_B T
\frac{ \mid { x}_{fi} \mid^2 }{ \hbar^2} 
\int_0^t dt_1 \int_0^{t_1} dt_2 \Gamma(t_1-t_2) \cos \Omega_{fi} (t_1-t_2)
\end{equation}
where ${ x}_{fi} = < E_f \mid { x}  \mid E_i > $ is the matrix
element of the coupling operator and $\Omega_{fi}= (E_f - E_i)/ \hbar$ . 
In the case of Ohmic dissipation with the static friction
coefficient $\gamma$,  the memory function is
proportional to the delta function, $\Gamma(t)= 2 \gamma
\delta(t)$, and Eq. (13) gives the rate  
\begin{equation}
\lambda_{fi} =
   \frac{2  }{\hbar^2} \mid {x}_{fi} \mid^2 \gamma k_B T ~~.
\end{equation}
 \\ \indent
This was obtained assuming a classical or
quasi-classical behavior of the environmental  degrees of
freedom. Therefore, it should provide a good approximation when the thermal 
energy $k_B T$ is greater than the transition energy, $\hbar 
\vert \Omega_{fi} \vert $. 
\\[.5cm]
{\bf 3. Thermally driven atom tunneling}
\\[.5cm] \indent
The atom dynamics in the STM junction will be
treated assuming that the CM motion is restricted to the X-axis, normal
to the surface, and the potential energy $V(x)$ has an asymmetric, 
double-well shape.  Without external bias, this potential is 
determined only by the binding  interaction  energy, and the
estimate provided  in ref. \cite{walk} can be well 
approximated by the fourth order polynomial
\begin{equation}
V_0(x)=C_0+C_1x +C_2 x^2+C_3 x^3+ C_4x^4 ~~, 
\end{equation}
with
$C_0 = 0.45$ meV, $C_1= 0.77$ meV/\AA, $C_2=-55.64$ meV/\AA$^2$,
$C_3=-11.59$ meV/\AA$^3$, $C_4=44.51$ meV/\AA$^4$.    
The isomeric  minimum of $V_0(x)$ is located at $x_0=-0.7$ \AA,
near surface, the barrier top at $x_b=0$, while the stable
minimum at $x_g=0.89$ \AA, near the tip.
 \\ \indent 
At the surface polarization by the voltage  $U$  with respect to
the tip, the dipole interaction energy changes the potential to
$V(x)= V_0(x)- Ed(x)$.  Here $E = U/2w$ is the junction electric field,
and $d(x)$ is the dipole moment of the Xe atom,
\begin{equation}
d(x)= Q_{eff} x + \mu_0  \{ \frac{1}{  0.3+0.7 
(w+x)^4/L^4 } - \frac{1}{ 0.3+0.7 (w-x)^4/L^4 } \}~~,
\end{equation}
where $Q_{eff} \sim 0.1 $ e  is the average effective charge of Xe
\cite{saenz},  $\mu_0=0.3$ Debye is the induced dipole moment at the surface,  
$w=2.2$ \AA, and $L=1.56$ \AA.  
\\ \indent 
If the temperature is zero and there is no dissipation, the atom dynamics 
during a voltage pulse $U(t)$ can be obtained by integrating the time-dependent
Schr\"odinger equation (TDSE) 
\begin{equation}
i \hbar
\frac{ \partial \psi (x,t) }{ \partial t} =
\lbrack - \frac{ \hbar^2}{2M} \frac{ \partial^2}{\partial x^2} + V_0(x) 
- \frac{U(t)}{2w} d(x) \rbrack \psi(x,t) ~~.
\end{equation}
Assuming that $U(0)=0$, the initial condition for integration,
$\psi(x,0) \equiv \psi_0(x)$, is represented by 
a Gaussian approximating the isomeric ground state of $V_0(x)$,
\begin{equation}
\psi_0(x) = (\frac{ c_0}{ \pi})^{1/4} e^{-c_0 (x-x_0)^2/2}
\end{equation}
where  $c_0= M \omega_0/ \hbar$, $M$ is the Xe mass,
and $ \omega_0 = \sqrt{ M^{-1} d^2 V_0/ dx^2 } \vert_{x_0}$. 
The integration of Eq. (18)  was performed numerically
in a spatial grid $\{ x_k \}$, $k=1,N$, by reduction to a Hamilton system of 
equations. Thus, if $u_k (t) \equiv Re ( \psi(x_k,t) )$ and 
$v_k (t) \equiv Im ( \psi(x_k,t ))$ denote
the real, respectively the imaginary part of the 
wave function $\psi(x,t)$  at the grid point $x_k$, then Eq. (18) becomes
\begin{equation}
2 \hbar \dot{u}_k =  \frac{ \partial {\cal H} }{ \partial v_k} ~~~~~~~~~~
2 \hbar \dot{v}_k = - \frac{ \partial {\cal H} }{\partial u_k}~~,
\end{equation}
with 
\begin{equation}
{\cal H} = \sum_{k=1}^N u_k  (T u)_k + v_k (T v)_k+ V(x_k) (u_k^2+v_k^2)~~,
\end{equation}
$$
(Ty)_k = - \frac{ \hbar^2}{2M dx^2} \lbrack \frac{y_{k+3}+ y_{k-3}}{90}
-3 \frac{y_{k+2}+y_{k-2}}{20} + 3 \frac{y_{k+1}+y_{k-1}}{2} - \frac{49}{18}
y_k \rbrack 
$$
The Hamiltonian system of Eq. (20)  was defined considering $N=321$ spatial 
grid points equally spaced by $dx=0.01$ \AA~ within the interval 
$[ x_{min}, x_{max} ]$= [$-1.2$ \AA, 2 \AA ]. For a fast integration 
was used the D02BAF routine of the NAG library \cite{nag},
with the time step $dt=6.58 \times 10^{-2} $ ps.  This time step
is by two orders of magnitude greater than the time step required
for the same accuracy by the leap-frog method. \\ \indent
The solution $\psi (x,t)$  can be used to calculate the time-evolution
of the localization probability in the stable well of $V_0(x)$, defined by
\begin{equation}
 \rho (t) = \int_{0}^{x_{max}} \vert \psi(x,t) \vert^2  dx ~~.
\end{equation}
The results obtained when $Q_{eff}=0$  for
a triangular voltage  pulse of 20 ns are presented in Fig.1(A), and for a 
trapezoidal pulse of 7 ns, in Fig.1(C).
The bias voltage during the pulse is represented in Fig.1(B) and (D),
respectively. The localization probability on the tip presented in
Fig.1(A) and 1(C),  increases during the pulse front
by sudden jumps at certain time moments $t^A_k$ and $t^C_k$. 
These time sequences are different, but the sequences $U^A_k$, $U^C_k$ of the
corresponding bias voltages, defined by $U^A_k=U(t^A_k)$ and $U^C_k=U(t^C_k)$
with $U(t)$ of Fig.1(B) and (D) are practically the same 
(e.g. $ -0.08$ V, $-0.164$ V, $-0.24$ V, $-0.32$ V, $-0.4$ V, $-0.48$ V, ...), 
equally spaced by $\sim 0.08$ V. These  jump voltages 
are  also very close to the values of $U$  known to ensure a resonant tunneling
of the Xe atom between the potential wells by QCO  \cite{gbc}.
Therefore, the jumps are explained by the crossing of the resonances
during the pulse front, here with arbitrarily chosen slope. 
\\ \indent
The features noticed above are present also when $Q_{eff}= 0.08$ e,
as indicated by the results obtained with a trapezoidal pulse of 
7 ns (Fig.2).  The increase of $\rho$ during the pulse front keeps the discontinuous 
character, but the jumps appear at different bias voltages,
(e.g. $-0.35$ V, $-0.39$ V, $-0.43$ V, $-0.47$ V, .... ), equally spaced
by $ \sim 0.04$ V. This result suggests that 
the effective charge represents an additional parameter which may shift  
the values and the spacing of the resonant voltages.  In Fig.2(A), 
$\rho$ becomes 1 during the voltage peak, but the residual value
after pulse is small. The propagation in time up to 20 ns of the 
non-stationary wave packet created by the pulse do not indicate further 
significant changes in the evolution of $\rho$.
 \\ \indent
The extension of the upper time limit of the numerical integration to
$\mu$s or ms  is not yet possible due to the large
amount of computer time required. Though, the present results
suggest that after the moment when the pulse vanishes, the  residual
probability  of localization on the tip is small.  \\ \indent
This behavior was obtained neglecting the dynamical effects
produced by the coupling between the atom and the surrounding electron gas. 
The average features of the atom-electron interaction are reflected 
by the jump in the junction conductivity at the atom switching, and
the potential energy term containing the effective charge, assumed above to 
be fixed. 
Though, the experiments on electromigration in metals show
that the effective charge is an average quantity which depends on temperature 
\cite{ver}. Therefore, it may have thermal fluctuations, and for a small value 
as $Q_{eff} \sim 0.1$ e, such effects could be important. \\ \indent
In the following, the fluctuating part of $Q_{eff}$ will be treated 
phenomenologically, assuming that it can be simulated by an additional dipole 
interaction in the Hamiltonian, having a structure close to $H_{coup}$ 
of  Eq. (1).     \\ \indent 
If the Xe atom adsorbed on the surface is thermally equilibrated, and the 
barrier crossing proceeds by quantum tunneling, then for short times 
compared to the recurrence time $\tau_R$ it could 
be possible to define the average transfer rate \cite{saenz}
\begin{equation}
\lambda = \frac{\sum_i e^{-E_i/k_BT^*} \lambda_i }{ \sum_i e^{-E_i/k_BT^*}}~,
\end{equation}
where $T^*$ is the effective temperature, $E_i$ are the energies of the 
isomeric levels  and $\lambda_i$ the corresponding tunneling rates. However, if
the system is not equilibrated, and at $t=0$ the CM wave function is a pure 
state $\psi_0$, then  the transfer rate should be defined by
\begin{equation}
\lambda (t)= \frac{ \dot{\rho}_{av}(t) }{1- \rho_{av}(t)} 
\end{equation}
 where $\rho_{av}(t) \equiv  << \rho >> (t)$ is the 
the average localization probability across the barrier ($x> x_b$), 
in the stable  well of  $V(x)$,
\begin{equation}
\rho_{av} (t)  = \int_{x_{b}}^{x_{max}} <x \mid {\cal D}_{av} (t) \mid x> dx~~.
\end{equation} 
According to Eq. (6), the matrix element $< x \mid {\cal D}_{av}(t)
\mid x> $ is given by
\begin{equation}
<x \mid {\cal D}_{av} (t) \mid x> =   \frac{1}{N_t} \sum_{r=1}^{N_t} 
\vert  <x  \mid \psi^r (t) > \vert^2~~.
\end{equation}
The probability amplitude  $< x \mid \psi^r(t) > \equiv  \psi^r (x,t)$ 
required here can be obtained by integrating the equation
\begin{equation}
i \hbar \partial_t  \psi^r(x,t) = \lbrack H_0 - x (\xi^r (t) - 
\gamma \frac{d <x>_r }{dt} ) \rbrack \psi^r(x,t)~~,
\end{equation} 
where $<x>_r \equiv < \psi^r \vert x \vert \psi^r >$ and
\begin{equation}
H_0= - \frac{ \hbar^2}{2M} \frac{ \partial^2}{\partial x^2} + V_0(x) 
- \frac{U}{2w} d(x) ~~. 
\end{equation}
The average in Eq. (26) was calculated using $N_t=100$ solutions
$\psi^r (t)$ of Eq. (27) at the environmental
temperature $T=4$ K and $U$ constant. Each solution was obtained
considering $\xi^r(t)$ at the moment $t_n=ndt$  of the form 
$
\xi^r (t_n) = R_n   \sqrt{  2 k_B T \gamma/dt} 
$
where $\{ R_n$, $n=1,2,3,.... \}$  is a sequence of Gaussian random numbers 
with $0$  mean and variance 1. This choice ensures the discrete form of
the FDT,  $<< \xi (t_j ) \xi (t_k ) >> = 2 k_B T \gamma \delta_{t_j t_k} / dt 
$. \\ \indent
The static effective charge parameter in  $d(x)$ is chosen $Q_{eff}=0$
and the friction coefficient is $\gamma =0.1$ $ \hbar /$ \AA$^2$.  
This corresponds to a partial damping rate $\gamma/ M \sim 5$ GHz, 
small compared to the total estimated rate of vibrational relaxation,
 $\sim 30$ GHz \cite{salam}.
\\ \indent
At $t=0$ the atom is supposed to be in a pure state localized near the 
surface,  and the initial condition for integration is
the ground state $\psi_0$ of the modified potential
\[ V_{mod}(x) = \left\{ \begin{array}{ll}
\mbox{ $ V_0 (x)- \frac{U}{2 w} d(x) $} & \mbox{ if $x \le x_b   $}
\\[0.2cm]
\mbox{$ V_0(x_b) - \frac{U}{2 w} d(x_b)
+ \alpha (x-x_b)^2 $} & \mbox{ if $ x>x_b $ ~~,} \\[0.2cm]
\end{array} \right. \]
obtained by replacing the stable well of $V(x)$  by a harmonic term. 
When $U$ decreases near the point when the barrier disappear, this 
procedure provides a better approximation for the isomeric ground state 
of $V(x)$ than the Gaussian wave packet of Eq. (19). 
\\ \indent
 The state $\psi_0(x)$ was obtained numerically by the Runge-Kutta integration 
of the   stationary Schr\"odinger equation
\begin{equation}
\lbrack - \frac{ \hbar^2}{2M} \frac{ d^2}{ d x^2} + V_{mod} (x)
\rbrack \psi_0 (x)= E_0 \psi_0 (x)~~,
\end{equation}
where  $V_{mod}$ is defined using $\alpha=0.1$. 
\\ \indent
The average localization probability on
the tip  $<< \rho>>$ and the average energy  $<<H_0>> \equiv Tr[H_0 
<< \rho>>]$ when $U=-0.8$ V are represented in Fig. 3(A) and (B)
by solid lines.  The time-dependence of $<< \rho>>$  
can be well approximated in this case by the  exponential  law
\begin{equation}
\rho_f (t)= 1 - e^{- \lambda_1 t}~~.
\end{equation}
The rate constant extracted by fit is $\lambda_1 = 210$ MHz, two orders 
of magnitude above the WKB tunneling rate, $\lambda_{WKB} = 6.2 $ MHz. 
The WKB rate characterize the non-dissipative tunneling for times smaller 
than the recurrence time, $\tau_R \sim 2$ ps. For longer times,
the integration of Eq. (27) with $\xi=0$, $\gamma=0$,  shows that
$\rho(t)$ has the anharmonic oscillatory evolution pictured in Fig.3(A) by 
dashed line. These oscillations contain high and low frequency components,
indicating that the initial wave packet is a superposition of 
several eigenstates of $H_0$. During the low frequency oscillation
the maximum attained by $\rho$ is $ \rho_m  \sim 0.5$ 10$^{-3}$, while the 
oscillation amplitude of the average  position, $<x>$, is $A_x = (<x>_{max} 
- <x>_{min})/2 \sim $ 10$^{-3}$ \AA. \\ \indent
The low frequency oscillation may be understood assuming that
the main components of  $\psi_0$  are two eigenstates $\psi_i$ and 
$\psi_f$ of $H_0$ close in energy, but localized in the isomeric, 
respectively in the stable well  of $V(x)$. 
In this case, the matrix element $ \vert x_{fi} \vert$ is well approximated
by $ A_x/ \sqrt{ \rho_m}= 0.045$  \AA.  With this value and $\gamma$, $T$ used 
to calculate $<< \rho >>$, Eq. (15) gives an inelastic tunneling rate 
$\lambda_{fi} = 209 $ MHz, very close to  $\lambda_1$ extracted by fit. 
\\ \indent
The two-state approximation becomes particularly suited
when the potential is tuned on a QCO resonance \cite{legg}.
At resonance $\psi_0$  is a linear superposition of two 
quasi-degenerate eigenstates of the Hamiltonian, $\psi_d$, $\psi_u$ each
being localized with probability 1/2 in either well. Their energies 
$E_d \approx E_u$ are such that the average level spacing $2 \pi \hbar
/ \tau_R$  is much greater than  $\Delta= \mid E_u - E_d \mid$, and the 
localization probability of $\psi_0$ "across the barrier" is
\begin{equation}
\rho_{QCO} (t) = \frac{1}{2}(1 - \cos \frac{2 \pi t }{ T } )~~,
\end{equation}
oscillating with the period $T= 2 \pi \hbar/ \Delta$, greater than 
$\tau_R$ . \\ \indent 
The QCO's of the STM Hamiltonian have an important role 
in the increase of the atom transfer probability during the pulse front, 
but they are also very sensitive to the environmental decoherence \cite{mg1}. 
To study the environmental effects on the atomic QCO the bias voltage
was fixed at the resonant value  $U_r= -1.141$ V when Xe oscillates
with the shortest period,  $T = 28.74$ ps.  Without environment coupling, 
($\gamma=0$), the localization probability near the tip for 
the initial wave packet  $\psi_0$ obtained from Eq. (29)  is pictured
in Fig.4(A) by dashed line. During this oscillation
$\rho_m \approx 1$, $A_x \approx 0.48$ \AA, and despite the anharmonicities, 
$\rho(t)$ is relatively close to $\rho_{QCO}(t)$. At the energy of
$\psi_0$ the WKB tunneling rate is $\sim 50$ GHz and $\tau_R \sim 3 $ ps.
\\ \indent
For an environmental temperature of $T=4$ K, the thermal energy is
$k_B T= 0.34$ meV, greater than the doublet splitting, $\Delta = 0.14$ meV,
and the decoherence effects should be important. 
The statistical average  $<<\rho>>$  defined by Eq. (25) was calculated using 
$N_t=100$ solutions of Eq. (27), and the result is pictured in Fig.4(A) by
solid line. The corresponding average of the energy,  $Tr[H_0 << \rho>>]$ is 
shown in  Fig.4(B).  \\ \indent
The average localization probability of Fig.4(A) has damped  oscillations  
which can be well  approximated by the formula
\begin{equation}
\rho_f (t) = 1 - u e^{- \lambda_1 t} - (1-u) e^{- \lambda_2 t}
\cos( \frac{2 \pi t }{ T + \delta } )~~.
\end{equation}
The parameters obtained by fit are 
$\lambda_1 = 3.6 $ GHz,  $\lambda_2 = 33$ GHz, $\delta = 1.6$ ps and 
$u=0.495$.  \\ \indent
In a two-state system ($\psi_u$, $\psi_d$) coupled 
to the environment the QCO amplitude is damped with the rate 
$\lambda_{ud}$ of the intradoublet transitions $\psi_u \leftrightarrow \psi_d$
\cite{bend}, and asymptotically the system arrives in the mixed state of 
maximum entropy \cite{mg2}.  In the present case $\psi_0$ is not a 
superposition of two states only, but Eq. (15) with the matrix element  
$\vert x_{ud} \vert \approx A_x = 0.48$ \AA~ gives
$\lambda_{ud}=24 $ GHz, close to $\lambda_2$ obtained by fit. 
\\[.5cm]
{\bf 4. Summary and conclusions} \\[.5cm] \indent
The atom transfer in the scanning tunneling microscope is a complex 
non-stationary phenomenon, reflecting a dynamical interplay between 
unitary evolution in Hilbert space and the environmental decoherence.
\\ \indent
A phenomenological description of a quantum system
interacting bilinearly with a classical heat bath of harmonic oscillators is 
provided by the modified Liouville equation presented in Sec. 1. This 
equation may be derived from a variational principle (Eq. (2)),
and has a Langevin form, containing stochastic and frictional terms. \\ \indent
The effects of a voltage pulse on the localization probability $\rho$ 
for a Xe atom prepared initially in a pure state localized on the STM 
surface was investigated by numerically integrating the TDSE (Eq.(18)).  
In these calculations the environmental interactions are neglected,
and the voltage pulse is assumed of symmetric triangular and trapezoidal 
shape. The results indicate a stepwise increase of $\rho$ at the
moments when the pulse front is near a resonant bias voltage for 
the isomeric ground state. The resonant values depend on the effective charge, 
being equally spaced by $\sim 0.08$ V  when $Q_{eff}=0$, and by $\sim 0.04 $ V
when $Q_{eff}=0.08$ e. However, the evolution of the non-stationary state 
created by the pulse do not indicate an asymptotic behavior
characteristic to exponential decay. \\ \indent
The atom dynamics at finite temperature was investigated in the 
frame of the stochastic, non-linear Liouville equation (Eq. (3)).  
The spectrum of the environmental noise was assumed to be flat (white noise), 
parameterized by the static friction coefficient $\gamma$. This 
has a value corresponding to weak damping, and the effective
temperature was considered the same as the environmental temperature,
$T=4$ K.    \\ \indent
The ensemble average of the energy increases during tunneling, (Fig.3(B) 
and 4(B)), but within the time interval of 100 ps investigated here the atom
is not thermalized.
The evolution of $<< \rho>>$ at resonance ($U=-1.141$ V),
consists of an incoherent superposition between damped QCO and exponential 
decay.  In the non-resonant case ($U=-0.8$ V)  the tunneling law
is close to exponential, with a rate proportional to the product $\gamma T$ 
(Eq. (15)). 
This result can provide a basis for understanding the current dependence of 
the atom transfer rate, because the friction coefficient as well
as the effective temperature should be functions of the electron tunneling 
current. 
 
\newpage
{\bf Figure Captions} \\[.5cm]

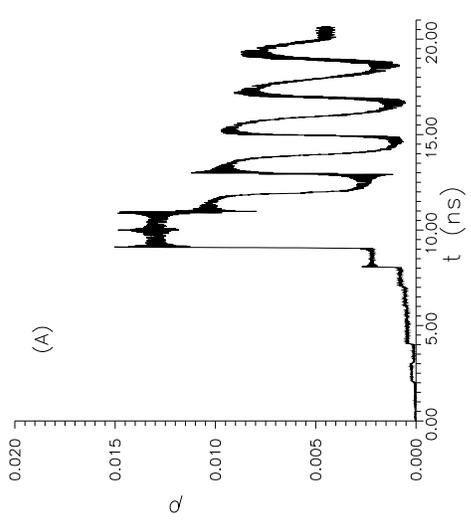
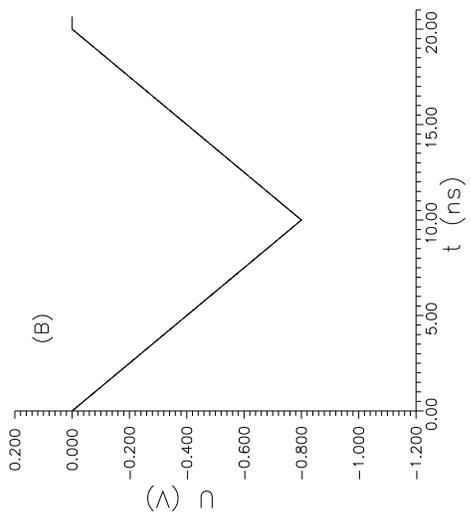
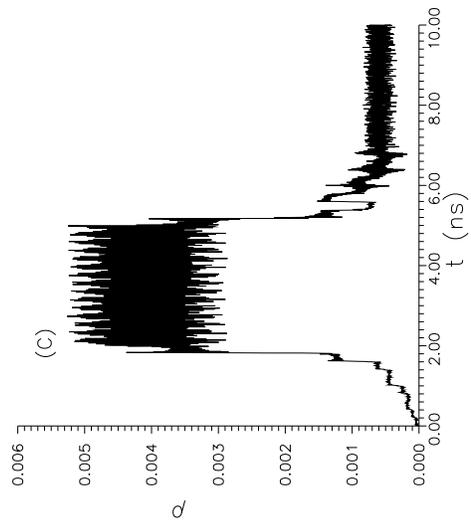
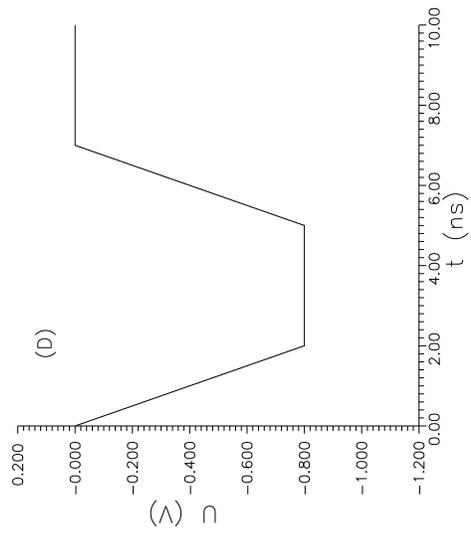
Fig. 1. Atom tunneling at $Q_{eff} =0$ without environment coupling.
$\rho$  and $U$ as a function of time for a triangular pulse of
20 ns (A),(B) and a trapezoidal voltage pulse of 7 ns (C),(D). \\

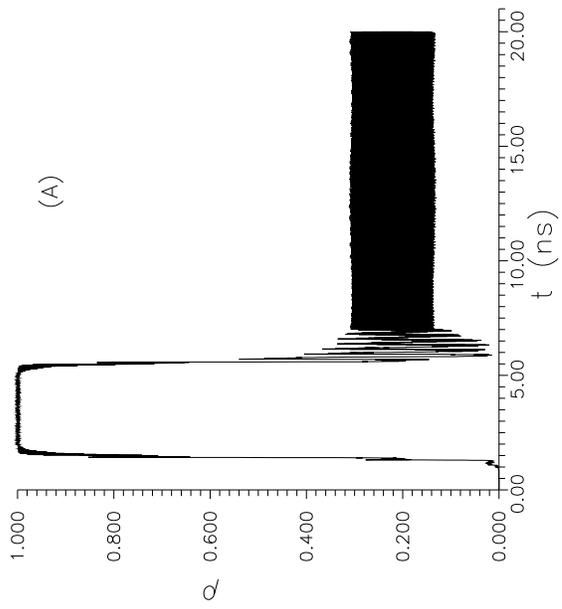
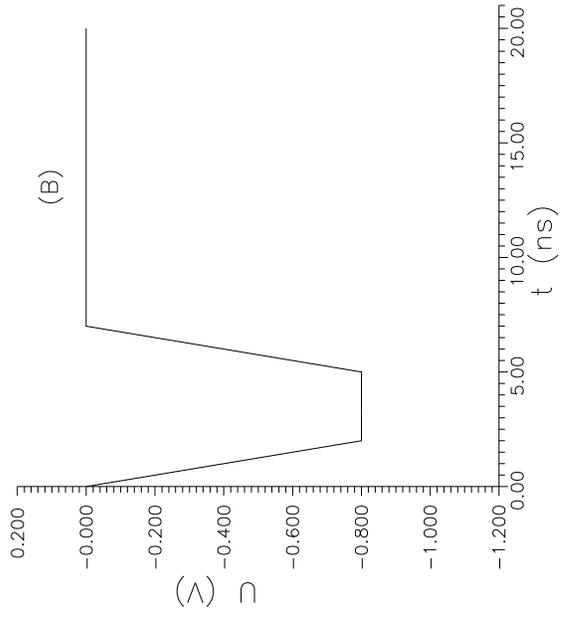
Fig. 2. Atom tunneling at $Q_{eff} =0.08$ e without environment coupling.
$\rho$ (A) and $U$ (B) as a function of time for a trapezoidal
voltage pulse of 7 ns. \\

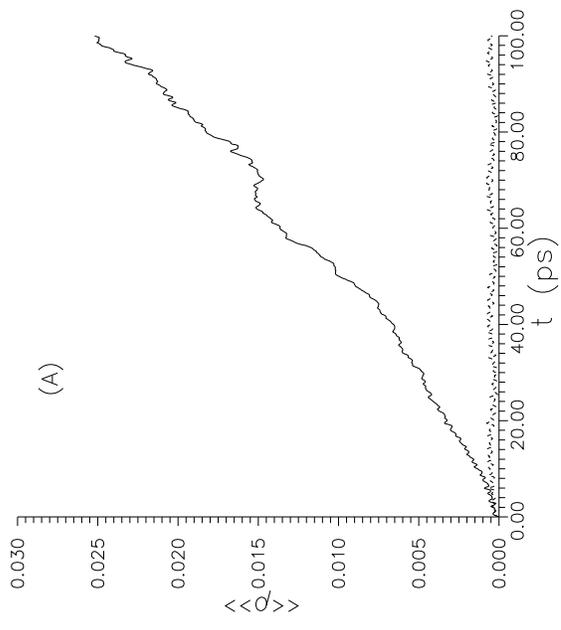
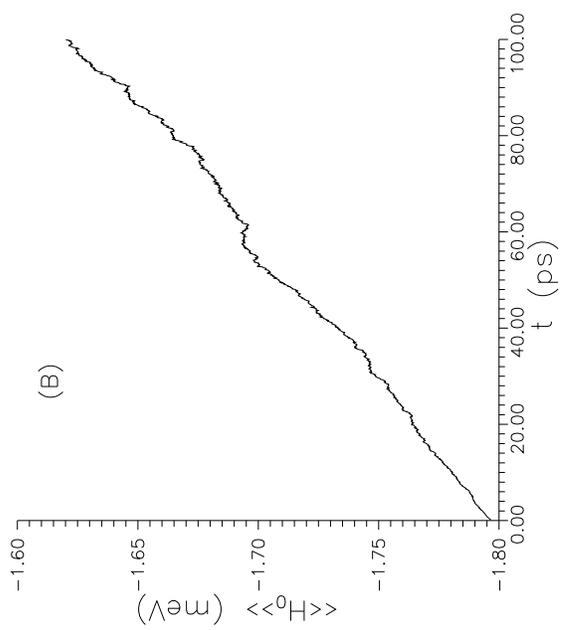
Fig. 3.  $<< \rho>> $ (A, solid) and $<<H_0>>$ (B) 
at $U=-0.8$ V for $T=4$ K, $\gamma=0.1 \hbar/ $ \AA$^2$, compared to
$\rho $ at $\gamma=0$ (A, dash), as a function of time.  \\

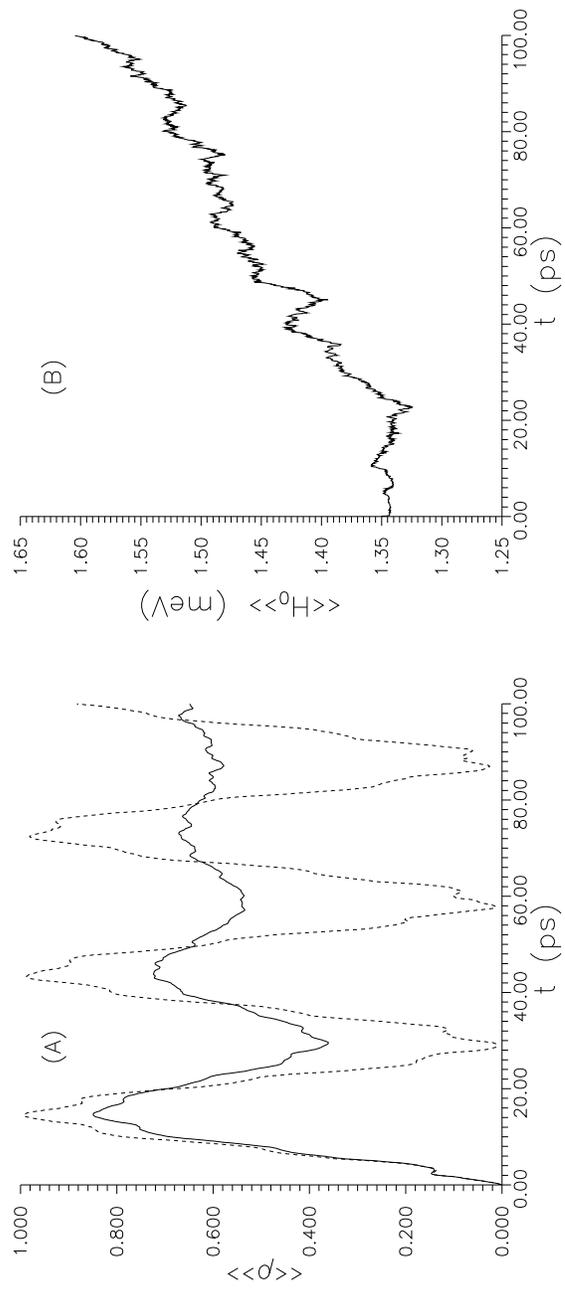
Fig. 4.  $<< \rho>> $ (A, solid) and $<<H_0>>$ (B) 
at $U=-1.141$ V, for $T=4$ K, $\gamma=0.1 \hbar/$ \AA$^2$, compared to 
$\rho$ at $\gamma=0$ (A, dash), as a function of time. 
\end{document}